\documentstyle[12pt]{article}
\begin{document}
\pagestyle{plain}
\newcommand{\be}{\begin{equation}}
\newcommand{\ee}{\end{equation}}
\newcommand{\bea}{\begin{eqnarray}}
\newcommand{\eea}{\end{eqnarray}}
\newcommand{\vp}{\varphi}
\newcommand{\pr}{\prime}
\newcommand{\sech} {{\rm sech}}
\newcommand{\cosech} {{\rm cosech}}
\newcommand{\psib} {\bar{\psi}}
\newcommand{\cosec} {{\rm cosec}}
\def\vs {\vskip .3 true cm}
\centerline { \bf{ The November $(J/\Psi)$ Revolution : Twenty-Five Years 
Later}}
\vs
\centerline { \bf{ Avinash Khare}}
\centerline {Institute of Physics, Sachivalaya Marg,}
\centerline {Bhubaneswar 751005, India.}
\centerline {email: khare@iopb.res.in}
\vs
{\bf Abstract}

Exactly twenty five years ago the world of high energy physics was set 
on fire by the discovery of a new particle with an unusually narrow
width at 3095 MeV, known popularly as the $J/\Psi$ revolution. This
discovery was very decisive in our understanding as well as
formulating the current picture regarding the basic constituents of
nature. I look back at the discovery, pointing out how unexpected,
dramatic and significant it was. 
\vfill
\eject
Exactly twenty five year ago, on November 10, 1974 two groups (one,
a MIT group doing experiment on the east coast at Brookhaven National
Laboratory, U.S.A. and the other a SLAC- Berkeley group doing
experiment on the west coast at Stanford Linear accelerator centre,
U.S.A)
simultaneously announced the discovery of a new particle at 3095 MeV
whose lifetime was about 1000 times longer than that of other particles 
of comparable mass. 
This announcement set on fire the world of high
energy physics and is now known in the physics community as the
November revolution. Like many revolutions, its meaning was not clear
at first. 
To appreciate the surprise, imagine if one suddenly discovers a new pyramid,
twice as heavy as the largest one known so far and yet a thousand times 
narrower and thus higher! 
The unprecedented large life time (or narrowness of the
peak) led to a host of theoretical speculations and vigorous
experimental activity. During the next one year, more than seven hundred papers
were written related to this discovery which was a record in physics
(if not in entire science) at that time. Subsequently, this record was 
broken after the discovery of high $T_c$ super-conductivity.

The $J/\Psi$ discovery electrified the community for many reasons including its
simultaneous discovery in two different laboratories and on two
entirely different type of machines. Its impact on the development
of high energy physics was tremendous and within two years of its
discovery, in 1976, the two men Samuel Ting and Burton Richter 
who led those two group
were awarded the Nobel prize in physics. Apart from Raman effect and 
parity violation, there are not many other discoveries which were
recognized that soon by the Nobel committee. The 
simultaneity has often led to speculation about how independent were
the two discoveries. Is it true that the work about the peak some how 
spread from Brookhaven to SLAC and played a role in the decision by
Richter and his group to set back the machine energy ? 

The plan of the
article is the following. I shall first briefly review the situation
as prevalent in high energy physics at the time of the discovery. Then 
I shall discuss in some detail the events leading to the
discovery. Finally, I shall discuss in brief the future developments
as well as the current status of the field.

{\bf Status of HEP in 1974 } 

The question of the basic constituents
of nature has attracted human mind since time memorial. {\it Atomos}
the Greek word of ``atom'' means indivisible, and it was thought for a 
long time that atoms were the ultimate, indivisible constituents of
matter. The discovery of electron at the end of the last century marks 
the end of the speculation era of about twenty five hundred years and
beginning of the realization that the atom is not indivisible or
elementary. In 1911, Rutherford showed that the atom consists of a
small dense nucleus surrounded by a cloud of electrons. Subsequently,
it was revealed that the nucleus consists of neutrons and
protons. During the fifties and the sixties, a large number of such
particles were detected. It was then suggested that perhaps all these
particles are also composite being made out of still smaller constituents
called quarks. Three kinds of quarks (u,d,s) and their anti-particles were
sufficient to describe all known hadrons at that time. In 1969, the deep
inelastic scattering experiments at SLAC (which were analogous to the 
Rutherford experiments of 1911) conclusively proved that 
proton, neutron 
and other particles are indeed composite being made out of quarks. 

The particles made out of quarks are called hadrons. These are the particles 
which interact through strong force. It is this strong force which binds 
protons and neutrons together in a nucleus. This force is also 
responsible for the rapid decay of many of the hadrons. 

In nature, there exist another kind of particles called leptons. By
1974, four such leptons were known i.e. electron, electron neutrino,
muon and muon neutrino ( and of course their anti-particles). The
leptons (unlike hadrons) do not experience strong force. Of course
electron and muon being electrically charged feel the
electromagnetic force (which is roughly hundred times weaker than the
strong force). The neutrinos have however no electric charge and hence 
they neither feel the strong nor the electromagnetic force, but
interact through weak interaction (it is weaker by several orders of
magnitude compared to the electromagnetic interaction). It
is worth pointing out these three together with gravitation were the
four basic interactions in nature at that time. Some people had
however already started wondering about the idea of unification of these 
forces. In particular, inspired by the unification of terrestrial and
celestial gravity by Newton and electricity and magnetism by
Maxwell, people were wondering if electromagnetic and weak
interactions could be unified into a single force. 
Such a unified model for leptons was
written down by Weinberg in 1967 but it did not receive much attention 
in the physics community till the work of 't Hooft who showed the
renormalizability of the model. The discovery of (weak) neutral current
in 1973-74 was a big boost to these ideas but initially several people 
including the experimental groups themselves were not sure of their
results so that it was jokingly termed as the discovery of ``alternating
neutral current '' and hence physics community had not grasped its full 
significance in 1974. I may add here that no Nobel prize has been
awarded to the discovery of neutral current.                                    

Thus in 1974 the common belief was that the basic constituents of
nature were three quarks and four leptons (and their anti-particles) 
and they all seemed point-like
objects. Among these, whereas the four leptons had been
experimentally seen in the laboratory as isolated particles, the
quarks always resided inside hadrons and no one was able to
isolate a single quark. This was one of the most puzzling aspect at
that time. 

{\bf Towards The Discovery}

I might point out that the situation in high energy physics was even
more confusing in 1970 when Ting wrote his proposal. Since 1965 Ting
had been working on the tests of quantum electrodynamic at high
momentum transfer. It may be noted that at small momentum transfer,
there is agreement between quantum electrodynamics 
and experiment about the anomalous magnetic moment of electron
to seven decimal places.
I know of no other
theory in physics (may be even in science?) in which there is such an
unprecedented agreement between theory and experiment.
From 1965 to 1969 Ting and his group observed the production of 
heavy photons (vector mesons) 
$\rho, \omega  $ and $\phi$ (whose masses were around 1 GeV) and
their subsequent decay to electron positron pair $(e^+e^-)$. One
obvious question was : how many heavy photons exist and what are their
masses and other properties ? Ting wanted to study this question but his 
proposal was rejected by both Fermilab and CERN. 
Finally he 
submitted a proposal to Brookhaven on January 11, 1972. He wanted to
look for heavy photons (vector mesons) 
through fixed target production experiments 
in which high energy protons will slam into a target
$$ p + p \rightarrow \ V^0 + X , \ 
X .... \ \  anything $$
and they will try to look for the heavy vector meson 
$V^{0}$ through its decay to $e^{+}e^{-}$ pair.
The interesting part of his proposal was his assertion ``contrary to
popular belief, the $e^+e^-$ storage ring is not the best place to
look for vector mesons. In the $e^+e^-$ storage ring the energy is
well defined. A systematic search for heavier mesons requires a
continuous variation and monitoring of the energy of the two colliding 
beams, a difficult task requiring almost infinite time. Storage ring is 
best suited to perform detailed studies of vector meson parameters
 once they have been found''. The subsequent events have confirmed this 
assessment and that is why hadron machine is popularly termed as a 
``discovery machine'' while $e^+e^-$ machine is meant for precision
studies.

Richter, on the other hand, was involved with $e^+e^-$ storage ring. He
was interested in understanding about the production of hadrons in
$e^+e^-$ collisions. In 1965, SLAC submitted a proposal to the US
atomic energy for such a machine with an energy of 3 GeV for each
beam. Funds were made available for this collider (SPEAR) only in 1970 
and the machine was built by April 72. The SPEAR group was primarily 
examining the ratio R which roughly speaking is the number of hadrons
divided by the number of muons produced in $e^+e^-$ collisions. More
precisely
$$R = {\sigma(e^+e^- \rightarrow \ hadrons)\over \sigma (e^+e^-
\rightarrow \mu^+\mu^-)}$$
where $\sigma$ denotes the cross-section. They wanted to study the
variation in the value of this ratio as the total energy is changed
from 2.4 GeV in steps of 200 MeV. From their preliminary study 
they found that this ratio R was 
rising from 2 to 6 as the total energy was increased from 2 to 5 GeV. Richter 
gave a talk about these results at the biannual Rochester conference
held in London in the summer of 1974. John Ellis also gave a talk at the same 
conference reviewing the production of hadrons in $e^+e^-$ collisions 
from different
models. He showed that depending on the model, this ratio could be
anything from 0.36 to $\infty$ (i.e. 0.36, 2/3, 2, 10/3, 4,...,$\infty)$.
The most widely accepted three quark model (with color) predicted R to 
be 2. Thus the situation appeared totally confusing as late as the
summer of 1974.

{\bf The Discovery}

Ting's proposal to Brookhaven 
got approved in May 1972 and was awarded a thousand
hours of beam time. It took the group almost 18 months to built the
detector which was enormous is every way: in size, intricacy,
sensitivity, and cost. The actual experiment started in April
1974. They first looked at $\phi$(1020) meson, the idea being, if
one plots the number of $e^+e^-$ pairs produced in this experiment 
as a function of total 
energy then one should see a broad peak whose maximum height is at
1020 MeV. One remarkable thing of his detector was that he could
measure the energy of the $e^+e^-$ pair with great accuracy. Of course
this made the detector very costly and he got lot of
criticism for making the detector needlessly accurate since no one at
that time thought that there could be a heavy vector meson with very
narrow width. On August 22, the team turned the detector to
energies between 2000 and 4000 MeV and took data for two weeks. Within 
couple of days of the data taking, two analysis teams independently
started to analyze the data (normally only one team analyzes the results but
Ting being a very careful man always had two) and both independently
realized that when they analyzed the number of events from 2875 to 
3225 MeV in gradation of 25 MeV, nearly all events were piled up at
3100 MeV i.e. {\it instead of a hill, they actually had a
needle}! And that was the big surprise as till then no subatomic
particle was known which had such a narrow width (i.e. such a large lifetime). 
It seemed to be 1000 times narrower than expected ! This is where the
personality of Ting came into  the picture. 

Several members of his group urged him to publish
the results immediately but he decided to doubly check the
results. But this was highly risky proposition since Ting knew 
that SPEAR could discover the peak in a day if
only they knew where the peak was ! On the other hand, Ting could
observe it in fixed target experiment only because of his obsessive
insistence on fine-tuning the detector. During this period 
 the MIT group members were making discrete enquiries about the
energy at which SPEAR was running and when they heard that it was
running between 4.5 GeV and 6 GeV, the group breathed freely!

Ting's frustrations increased when the machine restarted on October 2
but developed problems immediately. He then thought about announcing
the results during October 17-18 MIT festival to honour the retirement 
of Victor Weisskopf. However, he backed out at the last moment. On October
22, 
Ting
got back the machine to further recheck the data and  
one of his group member, Ulrich Becker gave a previously scheduled seminar 
at MIT where he disguised his very narrow peak by presenting the
number of events over a sufficiently wide energy range. However, it
did not fool Martin Deutsch who took Becker aside after the seminar
and asked him to publish the results immediately.  
On the same day, Mel 
Schwartz of Stanford stopped at Brookhaven to asses the progress of his
experiment. His assistant Jayashree Toraskar then told him about the
rumour out the bump at 3.1 GeV in Ting's experiment. Schwartz 
met Ting to get a confirmation about the rumour. The
conversion that followed had far reaching consequences. It is
therefore worth reproducing the conversation (as per Schwartz's
recollection).

Schwartz :  Sam, I hear you got a bump at 3.1.

Ting  :  \ No, absolutely not, Not only do I not have a bump, it's
absolutely flat.

Schwartz :  I will make you a bet. Ten dollars you get a bump.

Ting : \ Absolutely. I will bet.

Clearly, at least after this conversation, Ting should have announced the
discovery as he know very well that Schwartz was going back to
Stanford (where SLAC is situated) and once SPEAR hears the rumor about 3.1, 
they will just get it in a
day. By denying the rumour so flatly he in fact weakened his case and 
eventually he had to share the Nobel 
prize with Richter. There is no doubt in my mind
that if only he had been honest with Schwartz he would have got the full
credit.

On October 25, Deutsch again pressed the MIT group to publish the
results soon as otherwise SPEAR would get to it but still nothing
happened. Apparently, Ting now felt that there could be more than one
bump and he wanted to get credit for it too. 

On the west coast, on Saturday, November 9 
the SLAC group decided to stop their run between 4.5 and 6 GeV  
and instead go back to 3.1 GeV. 
Apparently, one member  of the group, Roy Schwitters felt that the
SPEAR group needed to write a paper 
on their experiment and hence he started to look at the data carefully.
While doing so, he noticed that there was something inconsistent in the data  
around 3.1 GeV. He talked with other group members including Gerson Goldhaber
and Richter and they agreed that indeed there was something odd and that one
should go back to 3.1 GeV. It may be 
noted that it is not easy to change energy just like that. One has also to
retune the beam and reset all the magnets. The  
official reason for going back is hardly convincing to 
say the least and it appears that the conversation of
Schwartz with Ting two weeks ago as well as other rumours floating around 
played some part in this decision. 
As expected,
within a day (i.e. on  Sunday, November 10) 
they had confirmed the existence of an unusually narrow
hadron at 3095 MeV. 

As the luck would have been, on the same day, Sunday November 10, Ting 
arrived  at
SLAC to attend a previously scheduled meeting of the SLAC Programme 
Advisory Committee. At the hotel, Ting got a frantic message from Deutsch 
who had heard that SPEAR had found something at 3.1 GeV. Ting was obviously
horrified and immediately telephoned Brookhaven and asked them to announce
the discovery right away and informed them that he will announce the discovery
at SLAC the next day. 
Ting also telephoned to the $e^+e^-$
machine at Frascati about the discovery and within two days they too
confirmed the existence of the particle. The three papers were
published back to back in the December 2 (1974) 
issue of Physical Review Letters. 

Why did Ting not published his data earlier 
when so many of his group members were urging him to do
so ? Was he ultra cautions or was he not very confident of the
results? Or was he greedy? Was he so naive to believe that his conversation
with Schwartz will not reach SLAC just because he had denied the rumour? 

While it is true that perhaps Ting should have got full credit for 
the discovery of $J/\psi$, the SPEAR group took over from that moment
onwards, doing all the precision studies related to the production and 
decay properties of $J/\psi$. Further, within ten days, the group
discovered another vector meson $\psi'(3695)$ MeV.

Soon after the announcement of the $J/\psi$ discovery, there followed a
host of theoretical speculations about what it is. The big question
was, why is $J/\psi$ so narrow and hence long lived? Some of the 
suggestions were
: it is an intermediate vector boson, Higgs boson, lightest colored
particle, lightest particle with a new quantum number called
paracharge, charm-anticharm $(c\overline c)$ quark bound state (
i.e. charmonium bound state). There followed a vigorous theoretical
activity trying to figure out the correct answer. In between, the SPEAR
group also discovered three more states through radiative transitions 
whose masses were between
those of $\psi'$ and $J/\psi$.  
Within about one year after the discovery it was clear
that $J/\psi$ was $c\overline c$ bound state so that the basic 
constituents of nature were were four
quarks and four leptons. The other states discovered at
SPEAR were also easily understood as the various states of the
charmonium $(c\overline c)$. Remarkably, it was shown that
the $c\overline c$ spectrum can be well understood within the framework
of non-relativistic quantum mechanics plus spin dependent
corrections. One last obstacle in this picture was the existence of
``Charmed mesons''. However even these were discovered by the middle
of 1976 and it convinced even the most die-hard skeptics about the validity
of the charm hypothesis.

Around this time followed two more rather unexpected discoveries,
namely those of $\tau$ lepton at 1786 MeV and $b\overline b$ bound
states where b is the fifth quark. I would say that these two were the
last two surprises and in the last twenty years 
we have not had any more {\it surprises} in High Energy Physics. 

It was clear by then that 
there are twelve basic constituent of nature i.e. six leptons
($\tau$-neutrino being the sixth lepton) and six quarks. Even though
only five quarks were known till then, the community was confident
that there must exist the sixth quark and such a quark (t-quark) was indeed
discovered at Fermilab in 1994.

{\bf The Present Picture}

As of today, 
the basic constituents of nature are six quarks and six leptons. The
strong interaction between quarks is due to their colour degree of
freedom and the corresponding  gauge quanta are called gluons and this
theory is known as quantum chromodynamics. On the other hand, the
electromagnetic and weak interaction between quarks and leptons is
given by a unified electroweak theory $SU(2)_L\otimes U(1)_Y$. By now 
all its predictions have been experimentally verified except for the
prediction of a neutral Higgs boson. 
A Large Hadron Collider (LHC)
machine is being built at CERN specifically to look for this particle
and it is expected that this issue will be resolved by the year 2007. 

It must be made clear that there are several basic questions that have
not been answered by the above (so called) ``Standard Model''. 
For example, the standard model has several arbitrary parameters. Besides,
the origin of fermion masses is unclear. Further,
there is only a partial unification of the basic forces. In recent years
a truely unified theory called the {\it superstring theory} has been
proposed which unifies all the four interactions. One remarkable break
from the past is that here the basic constituents of nature are not
particles at all! Rather the basic object is a string of length 
$10^{-33}$ cm. The quarks, leptons and the gauge bosons are merely the 
different modes of vibration of the string. The unification ideas have 
brought closer the seemingly contrasting worlds of the smallest and the 
largest. In particular, these ideas hold the promise to explain how the 
universe evolved a very short time after the big bang. Another possibility is
that the quarks and leptons are themselves composed of more elementary 
objects. It must be made very clear here that 
unfortunately, so far  
we have no experimental evidence for any of the ideas   
beyond the standard model. 

{\bf Suggested Reading}
 
{1.} R.P. Crease and C.C. Mann, {\it The Second Creation}, (Macmillan 
Publishing Company, U.S.A., 1986).  

{2.} Nobel Lectures by S. Ting and B. Richter (1976).

\end{document}